\documentclass[11pt]{article}
\usepackage{amsfonts}
\usepackage{amssymb}
\usepackage{amscd}

\def\a{\alpha}
\def\b{\beta}

\def\de{\delta}
\def\e{\epsilon}

\def\ve{\varepsilon}

\def\s{\sigma}
\def\th{\theta}
\def\z{\zeta}

\newcommand{\cl}{{\mathcal L}}

\newcommand{\C}{\mathbb C}
\newcommand{\R}{\mathbb R}

\renewcommand{\Im}{{\mathop{\mbox{Im}}\nolimits\,}}

\def\pa{\partial}
\def\dt#1{{\buildrel {\hbox{\bf .}} \over {#1}}}  
\def\ddt#1{{\buildrel {\hbox{\bf ..}}\over {#1}}}
\def\ad{{\dt{\alpha}}}
\def\bd{{\dt{\beta}}}
\def\gd{{\dt{\gamma}}}
\def\add{{\ddt{\alpha}}}
\def\bdd{{\ddt{\beta}}}

\def\dz{{\dt{0}}}
\def\ddz{{\ddt{0}}}
\def\ddo{{\ddt{1}}}

\def\sfrac#1#2{{\textstyle\frac#1#2}}

\def\beq{\begin{equation}}
\def\eeq{\end{equation}}
\def\beqx{\begin{displaymath}} 
\def\eeqx{\end{displaymath}}
\def\beql{\arraycolsep .1em \begin{eqnarray}}
\def\eeql{\end{eqnarray}}

\def\gl#1{(\ref{#1})}

\def\theequation{\thesection.\arabic{equation}}

\catcode`@=11
\@addtoreset{equation}{section}
\@addtoreset{equation}{subsection}
\def\theequation{\ifnum\value{section}=0 \arabic{equation}\ignorespaces
\else \ifnum\value{section}=-1 A.\arabic{equation}\ignorespaces
\else \ifnum\value{subsection}=0 \thesection.\arabic{equation}\ignorespaces
\else \thesection.\arabic{subsection}.\arabic{equation}\ignorespaces
                           \fi
                      \fi
                 \fi}
\catcode`@=12

\def\be{\begin{equation}}
\def\ee{\end{equation}}
\def\bea{\begin{eqnarray}}
\def\eea{\end{eqnarray}}

\def\N2{$N{=}2$}

\begin{document}
\begin{titlepage}
\begin{flushright}
hep-th/0009144\\
ITP--UH--15/00\\
\end{flushright}

\vskip 2.0cm

\begin{center}

{\Large\bf  On the Integrability of Covariant Field Theory\\
\vskip 0.2cm
for Open N=2 Strings  }

\vspace{14mm}

{\large Olaf Lechtenfeld \ and \ Alexander D. Popov~$^*$}
\\[5mm]
{\em Institut f\"ur Theoretische Physik  \\
Universit\"at Hannover \\
Appelstra\ss{}e 2, 30167 Hannover, Germany }\\
{Email: lechtenf, popov@itp.uni-hannover.de}

\end{center}

\vspace{2cm}

\begin{abstract}

We discuss the integrability of the Berkovits-Siegel open
string field equations and derive an infinite set of their non-local
(solution-generating) symmetries. The string field equations are
embedded in an infinite system of overdetermined equations
(BS hierarchy) associated with hidden string symmetries.
The latter enforce the vanishing of most scattering amplitudes for
the open $N{=}2$ string.

\end{abstract}

\vfill

\textwidth 6.5truein
\hrule width 5.cm
\vskip.1in

{\small
\noindent ${}^*$
On leave from Bogoliubov Laboratory of Theoretical Physics, JINR,
Dubna, Russia}

\end{titlepage}


\section{Introduction}

Although string field theory has not yet led to a non-perturbative
and background-independent formulation of string theory, it is a
useful tool to illuminate the relation between string and
particle dynamics. Besides popular issues like tachyon condensation
and non-commutative products, string field theory teaches us how
to extend gauge theory and gravity to the string realm, by including
the dynamics of an infinite tower of massive excitations.

\medskip

Such an extension would appear to be rather trivial for {\it self-dual\/}
field theories in $2{+}2$ dimensions: The corresponding critical string theory
has \N2 world-sheet supersymmetry but lacks any massive physical excitations!
Its massless ground state describes gluons and gravitons of same helicity only,
but in a light-cone gauge-fixed formulation~\cite{OV}.
Yet, this system shows $SO(2,2)$ Lorentz covariance in the sense that
its field equations, the self-duality conditions\footnote{
We use $4D$ real Weyl spinor notation for space-time indices,
$\a\leftrightarrow SL(2,\R)$ and $\ad\leftrightarrow SL(2,\R)'$,
with $\a=0,1$ and $\ad=\dz,\dt{1}$.}
\be
\label{coveqs}
R_{\a\b}\ =\ 0 \qquad\qquad\qquad F_{\a\b}\ =\ 0 \quad,
\ee
can be expressed covariantly while there is no Lorentz-invariant action for
them alone. This state of affairs makes it awkward to extend the theory
off-shell, but it can be remedied by including in the Lagrangian density~$\cl$
the other helicity degree of freedom
as dynamical Lagrange multiplier~\cite{CS},
\be
\label{covlag}
\cl\ =\ {\rm Tr} ~B^{\a\b} F_{\a\b}
\qquad\qquad\qquad
\cl\ =\ {\rm Tr} ~\rho^{\a\b} R_{\a\b}
\ee
(all fields are Lie algebra-valued).
One may wonder if `integrating in' such multiplier fields is also possible
on the string theory side and may lead to a Lorentz-covariant
\N2 {\it string field theory\/}.

\medskip

Indeed, Berkovits and Siegel~\cite{BS} have proposed such a covariant
string field theory for the open \N2 string, based on an earlier
non-covariant version of Berkovits~\cite{B}.
Their action differs from the standard Chern-Simons-like action~\cite{Wi}
in two ways. First, it contains not a single but {\it two\/} string fields,
$\Phi$ and $\tilde{\Phi}$, where $\Phi$ contains the prepotential for the
self-dual gauge field~$A_{\a\ad}$, and $\tilde{\Phi}$ functions as a
Lagrange multiplier string field multiplying the $\Phi$ equation of motion.
Second, in analogy with self-dual Yang-Mills~\cite{CS},
$\Phi$ and $\tilde{\Phi}$ are interpreted as a prepotential for a covariant
{\it string\/} field~$A_\a$ and a component of a Lagrange multiplier
{\it string\/} field~$B_{\a\b}$ for the self-dual field strength
$F_{\a\b}(A)$, respectively.
Note that these covariant string fields carry space-time indices,
but transform only under {\it one\/} of the two $SL(2,\R)$ pieces of
the $SO(2,2)$ Lorentz group, because the space-time derivative~$\pa_{\a\ad}$
of the field theory gets replaced by the action of two BRST operators~$G_\a$
for the \N2 string, obtained from a ``small'' $N{=}4$ superconformal algebra
after twisting. The covariant \N2 string field action
\be
\label{covsft}
S\ =\ \int {\rm Tr}~ B^{\a\b} * F_{\a\b}(A)
\ee
is formulated using Witten's string field product and integral~\cite{Wi}
(all fields are {\it string\/} fields now).
In the non-covariant Yang or Leznov gauges,
the string field equations of motion based on Eq.\gl{covsft},
\be
\label{coveq}
0\ =\ F_{\a\b}\ =\ \{ \nabla_\a , \nabla_\b \}
\qquad\quad{\rm with}\qquad
\nabla_\a\ =\ G_\a + A_\a \quad,
\ee
reduce to (Chern-Simons-like) Yang-type~\cite{Y}
or (quadratic) Leznov-type~\cite{L} equations for the string field~$\Phi$
(viz. $\pa_{\a\ad}\to G_\a$),
but possess residual gauge invariances due to the non-empty kernels of~$G_\a$.
Also, in contrast to the standard NSR $N{=}1$ string field theory~\cite{Wi},
the action~\gl{covsft} and its gauge-fixed variants do not suffer from
tree-level divergences due to colliding picture-raising operators~\cite{Wendt}.
This problem, however, can be avoided also in the NSR formulation by
properly distributing left- and right-moving picture-raisers~\cite{LeSa,talk}.

\medskip

In this paper, we generalize to the string level the twistor construction
\cite{W,WW} of self-dual gauge fields.
Namely, using an extra parameter $\z\in\C P^1$, we introduce a
free string field encoding all information about the solutions of the
Berkovits-Siegel (BS) string field equations~\gl{coveq}.
The two operators $G_\a$ are connected with target space-time coordinates.
To the (other) coordinates on the moduli space of self-dual gauge fields
we correspond new BRST-like operators. A hierarchy of string field equations
emerges, containing the BS string field equations~\gl{coveq} as well as
equations governing their symmetries. In this way, the hidden non-local
symmetries of the self-dual Yang-Mills (SDYM) equations get elevated
to the string field theory level.

\section{Covariant field theory for open N=2 strings}

{}From the world-sheet point of view, critical open \N2 strings
in flat Kleinian space $\R^{2,2}$
are a theory of \N2 supergravity on a $1{+}1$ dimensional
(pseudo) Riemann surface with boundaries,
coupled to two chiral \N2 massless matter multiplets $(X,\psi)$.
The latter's components are complex scalars (the four string coordinates)
and $SO(1,1)$ Dirac spinors (their four NSR partners).
The world-sheet action enjoys local
\N2 super coordinate and Weyl invariance on the world-sheet,
as well as global $U(1,1)$ target space-time symmetry.
In the superconformal gauge, the associated left-moving constraints
\beqx
T\ =\ \pa_z X^{\a\bd}\pa_z X_{\a\bd}
     +\psi^{\ad\bdd}\pa_z\psi_{\ad\bdd}\quad,
\eeqx
\be \label{0}
G^{0\ddo}\ =\ \psi^{\gd\ddo} \pa_z X^0\!_\gd \quad, \qquad
G^{1\ddz}\ =\ \psi^{\gd\ddz} \pa_z X^1\!_\gd \quad,
\ee
\beqx
J^{\ddz\ddo}\ =\ \psi^{\gd\ddz} \psi_\gd^{\ \ddo}
\eeqx
form a $c{=}6$ \N2 superconformal algebra,
where $\add\leftrightarrow SL(2,\R)''$ is the world-sheet internal index,
$\add=\ddz,\ddo$ associated with the local $U(1)$ R symmetry.
In the NSR formulation, the standard quantization procedure proceeds
by introducing ghost systems, extending the contraints to a $c{=}0$ algebra,
writing the BRST operator, and computing its (relative) cohomology
for various ghost and picture numbers.

Alternatively, Berkovits and Vafa~\cite{BV} developed a ghost-free formulation
of the \N2 string by making use of the extension of~\gl{0}
to the ``small'' $N{=}4$ superconformal algebra
\beqx
T\ =\ \pa_z X^{\a\bd}\pa_z X_{\a\bd}
     +\psi^{\ad\bdd}\pa_z\psi_{\ad\bdd}\quad,
\eeqx
\be \label{1}
G^{\a\bdd}\ =\ \psi^{\gd\bdd} \pa_z X^{\a}\!_{\gd}\quad,
\ee
\beqx
J^{\add\bdd}\ =\ \psi^{\gd\add}\psi_{\gd}^{\ \bdd}\quad.
\eeqx
After twisting this algebra,
$G^{\a\ddz}$ become two fermionic {\it spin-one\/} generators
which subsequently serve as BRST-like currents.
In this paper, we denote them by $G_\a :=\e_{\a\b}G^{\b\ddz}$.

\medskip

{}Following Berkovits and Siegel~\cite{BS},
we introduce two Lie algebra-valued fermionic string fields $A_\a[X,\psi]$,
where the arguments $X^{\a\ad}(\sigma)$ and $\psi^{\ad\add}(\sigma)$
with $\sigma{\in}[0,\pi]$ denote the open string configuration.
(Note that the Lagrange multiplier string fields, mentioned in
the Introduction, reside in another multiplet, $B_{\a\b}[X,\psi]$.)
Although we suppress it in our notation, string fields are always multiplied
using Witten's star product~\cite{Wi}.
The three BS~string equations of motion~\gl{coveq} read\footnote{
Since the star product is not (graded) commutative,
field squares do not vanish.}
\be\label{2}
G_1A_1+A_1^2=0\ , \quad
G_0A_1+G_1A_0+A_0A_1+A_1A_0=0\ ,\quad
G_0A_0+A_0^2=0\ ,
\ee
and admit the following two gauges for $A_\a$:
\bea\label{3}
A_0&=&0 \qquad{\rm and}\qquad A_1\ =\ e^{-\Phi}G_1e^{\Phi} \quad ,\\[6pt]
\label{4}
A_0&=&0 \qquad{\rm and}\qquad A_1\ =\ G_0{\Psi} \quad ,
\eea
where $\Phi$ and $\Psi$ are bosonic string fields.
The reality conditions on these string fields are \cite{BS}:
\be\label{2.6}
\Phi^\dagger [X(\s),\psi(\s)]\ =\ \Phi[X(\pi-\s),\psi(\pi-\s)]\qquad,\quad
\Psi^\dagger [X(\s),\psi(\s)]\ =\ \Psi[X(\pi-\s),\psi(\pi-\s)] \quad.
\ee
The gauges \gl{3} and \gl{4}
are analogues of the Yang~\cite{Y} and Leznov~\cite{L}
gauges in the SDYM theory.
For the Yang gauge \gl{3}, Eqs.\gl{2} are reduced to
\be\label{5}
G_0(e^{-\Phi}G_1e^{\Phi})\ =\ 0\quad ,
\ee
whereas for the Leznov gauge~\gl{4}, they simplify to
\be\label{6}
G_1G_0\Psi + G_0\Psi G_0\Psi\ =\ 0\quad .
\ee
The Yang-type Eq.\gl{5} as well as the Leznov-type Eq.\gl{6} are
non-linear generalization of the ``wave'' equation $G_1G_0\Psi=0$.

\medskip

The gauge symmetries of Eqs.\gl{2}, \gl{5}, and \gl{6} are described
in~\cite{BS}.  Concretely,
the covariant equations~\gl{2} are invariant under the transformations
\be\label{7}
A_\a\ \mapsto\ \tilde A_\a\ =\ e^K (G_\a + A_\a) e^{-K}\quad,
\ee
where $K$ is arbitrary.
For the gauge-fixed equations \gl{5} and~\gl{6},
this translates to invariances under
\be\label{8}
e^{\Phi}\ \mapsto\ e^{\tilde\Phi}\ =\ e^{\Phi}e^{-N}
\qquad{\rm with}\quad
N=G_0\Xi  \quad  ,
\ee
and
\be\label{9}
\de\Psi\ =\ (G_1\Xi  + [G_0\Xi , \Psi ]) \quad,
\ee
respectively, where $\Xi$ is an unconstrained string field.

\section{Integrability of the BS string field equations}

In this section we obtain the BS string field equations \gl{2}
as the {\it compatibility conditions\/} of some linear equations and
give a general recipe for solving them.
Consider the extended complex plane $\C\cup\{\infty\}=\C P^1$
with a coordinate $\zeta$ and introduce on it two coordinate patches,
\be\label{10}
\bar H^2_+\ =\ \{\zeta\in\C\cup\{\infty\}: \Im\zeta\ge 0\}
\qquad {\rm and} \qquad
\bar H^2_-\ =\ \{\zeta\in\C\cup\{\infty\}: \Im\zeta\le 0\}\quad ,
\ee
with the overlap
\be\label{11}
S^1\ =\ \R P^1\ =\ \bar H^2_+\cap\bar H^2_-\ =\
\{\zeta\in\C\cup\{\infty\}: \Im\zeta = 0\}\quad .
\ee
Using $G_\a$, $A_\a$, and $\zeta$, we define the following linear
equation
\be\label{12}
(G_1+\zeta G_0 + A_1 + \zeta A_0)\Psi_+\ =\ 0\quad ,
\ee
where $\Psi_+=\exp (\Phi_+)$, and $\Phi_+$ is a string field
depending not only on $X(\sigma )$ and $\psi (\sigma )$
but also on the parameter $\zeta\in\bar H^2_+$. If $A_\a$ is given,
then \gl{12} is an equation for the field $\Psi_+$. Solutions
$\Psi_+$ of this linear equation exist (and are functionals of $A_\a$)
if the equation
$$
(G_1+\zeta G_0 + A_1 + \zeta A_0)^2\ =\ 0\qquad\Leftrightarrow
$$
\be\label{13}
(G_1A_1+A_1^2)+\zeta (G_1 A_0 + G_0 A_1+ A_1 A_0+A_0 A_1)+\zeta^2
(G_0A_0+A_0^2)\ =\ 0
\ee
is satisfied for  any $\zeta\in\bar H^2_+$. This statement is equivalent
to the three BS equations \gl{2}. Moreover, Eqs.\gl{2}
are also the integrability condition for the linear equation
\be\label{14}
(G_1+\zeta G_0 + A_1 + \zeta A_0)\Psi_-\ =\ 0\quad ,
\ee
where $\Psi_-=\exp (\Phi_-)$, and $\Phi_-[X(\sigma),\psi(\sigma),\zeta]$
is a string field depending also on $\zeta\in\bar H^2_-$.

\medskip

Two remarks are in order.
First, finding $\Psi_\pm$ for given $A_\a$ is a
``direct transform'' $A_\a\to\Psi_\pm$ in the terminology of
integrable systems, and finding $A_\a$ for given $\Psi_\pm$ is
an ``inverse transform'' $\Psi_\pm\to A_\a$.
Second, Eq.\gl{13} can be cast in Chern-Simons form
\be\label{15}
Q_0A+A^2\ =\ 0
\ee
by introducing
\be\label{16}
Q_0\ :=\ G_1+\zeta G_0
\qquad {\rm and} \qquad
A\ :=\ A_1+\zeta A_0 \quad .
\ee
Hence, the BS equations follow from the Chern-Simons-like equation \gl{15}.
Fields killed by the operator $Q_0$ will be called chiral.

\medskip

Notice that the gauges \gl{3} and \gl{4} are connected with the
following respective choices of asymptotic behavior for $\Psi_\pm$
(cf. \cite{IvLe})
\bea\label{17}
\Psi_\pm &=& e^{-\Phi}+O(\zeta)\qquad\qquad\qquad
{\mbox {for}}\quad \zeta\to 0\quad , \\[6pt]
\label{18}
\Psi_\pm &=& 1+\zeta^{-1}{\Psi}+O(\zeta^{-2})\qquad\ \;
{\mbox {for}}\quad \zeta\to\infty\quad .
\eea
Indeed, inserting \gl{17} or \gl{18} into Eqs.\gl{12} and \gl{14}
yields formula \gl{3} or \gl{4}.
Recall that the string fields $\Phi$ in \gl{3}, $\Psi$ in \gl{4},
and $\Phi_\pm$ in \gl{12} and \gl{14} carry Chan-Paton factors
which are suppressed throughout this paper.

\medskip

Now let us introduce the string field
\be\label{19}
\Upsilon_{+-}\ :=\ \Psi_+^{-1}\Psi_-\ =\ e^{-\Phi_+}e^{\Phi_-}
\ee
defined on the overlap $\bar H^2_+\cap\bar H^2_-=S^1$,
so that it also depends on $\zeta\in S^1$.
{}From Eqs.\gl{12} and \gl{14} it follows
that
\be\label{20}
Q_0\Upsilon_{+-}\ \equiv\ (G_1+\zeta G_0)\Upsilon_{+-}\ =\ 0\quad ,
\ee
i.e. $\Upsilon_{+-}=: e^{\Phi_{+-}}$ is a {\it free} chiral string field
depending on an extra parameter $\zeta =\cot\sfrac\th2,\ 0\le\theta\le 2\pi$.
Thus, constructing solutions of the BS equations
\gl{2} is equivalent to finding a field $\Upsilon_{+-}$
satisfying \gl{20} and splitting $\Upsilon_{+-}=\Psi_+^{-1}\Psi_-$.
Then $A_\a$ can be obtained from $\Psi_\pm$ by formulae \gl{12}, \gl{14}
or directly from  \gl{17}, \gl{18} and \gl{3}, \gl{4}. Notice that the
zero-mode sector of $\Upsilon_{+-}$ contains a transition matrix of
a holomorphic bundle over the twistor space \cite {W,WW} which encodes
all information about self-dual gauge fields.

\section{Symmetries and a BS hierarchy}

We now analyze Grassmann-type symmetries generalizing the known symmetries
\cite{BS}   $\de_\a\Psi = G_\a\Psi$ of the BS equations \gl{2}
in the Leznov gauge \gl{4}. In other words,
we investigate infinitesimal  deformations $\Psi\mapsto \Psi +\ve\de\Psi$
of solutions to the Leznov-type Eq.\gl{6}, where $\ve$ is a constant
Grassmann parameter.
Clearly, $\de\!:\Psi\mapsto\de\Psi$ deserves to be called an infinitesimal
symmetry transformation if with $\Psi$ also $\Psi{+}\ve\de\Psi$ satisfies
Eq.\gl{6} to first order in $\de\Psi$, from which it follows that
\be\label{21}
G_1G_0\de\Psi + G_0\Psi G_0\de\Psi - G_0\de\Psi G_0\Psi\ =\ 0\quad .
\ee
To find infinitesimal symmetries of Eq.\gl{6} means to find solutions
$\de\Psi$ of Eq.\gl{21} for any given solution $\Psi$ of Eq.\gl{6}.
Analogously one may consider symmetries of Eq.\gl{5}.

\medskip

The BS equations \gl{2} and their reduced forms \gl{5} and \gl{6}
can be embedded in an infinite system of overdetermined equations,
in the sense that every solution to the BS equations can be extended to
a simultaneous solution of this infinite system. The latter will be called
the BS hierarchy of equations and describes, in particular,
the symmetries of the BS equations which we shall derive.

\medskip

We recall that the zero-mode sector of Eq.\gl{6} is precisely the
Leznov field equation, describing SDYM in the Leznov gauge~\cite{BS}.
{}From SDYM theory we know
that the solution of the Leznov field equation depends not only
on the space-time coordinates $x^{0\ad}$ and $x^{1\ad}$ but
also on coordinates $x^{2\ad},\ldots,x^{2J\,\ad}$  on the moduli
space of self-dual gauge fields, where $2J$ is any positive
integer or infinity. In fact, this dependence may be considered as a
manifest one (see \cite{IvLe} and references therein).
When lifting the solution of the Leznov field equation to the
Leznov {\it string\/} field~$\Psi[X,\psi]$ solving~\gl{6},
it is therefore natural to promote not only the space-time coordinates
but also these moduli coordinates to the string level,
\be
X^{m\ad}(\s,\tau)\ =\ x^{m\ad} + p^{m\ad}\tau + {\rm oscillations}
\qquad{\rm for}\quad m=0,\ldots,2J \quad,
\ee
and consider all string fields as fields
depending on $\psi^{\ad\bdd}(\s)$ and $X^{m\ad}(\s)$.
Effectively, we have extended the range of the space-time $SL(2,\R)$
spinor indices $\a=0,1$ to $m=0,\ldots,2J$.
In this sense, the $N{=}2$ string lives in a $2(2J{+}1)$ dimensional
target space-time.
In the first-quantized approach,
one may consider the operator product expansions
\be\label{22}
X^{m\ad}(z,\bar z)\ X^{n\bd}(w,\bar w)\ \sim\
g^{m\ad , n\bd}\,\ln |z-w|^2 \quad,
\ee
where $g_{m\ad , n\bd}$ is a suitable non-degenerate metric on the 
$2(2J{+}1)$ dimensional target space-time, which,
upon restriction to the Kleinian 4$D$ space-time,
reduces to the flat metric $g_{m\ad,n\bd}=\epsilon_{mn}\epsilon_{\ad\bd}$ 
with $m,n=0,1$.
We shall also use coordinates
\be\label{4.4}
X_{m\ad}\ :=\ g_{m\ad , n\bd} \, X^{n\bd}\quad .
\ee

\medskip

Generalizing $G_\a\!^\bdd$, we introduce the operators
\be\label{24}
G_m\!^{\bdd}\ =\ \psi^{\dt\gamma\bdd}\pa_zX_{m\dt\gamma}\quad .
\ee
After twisting one obtains the BRST-like fermionic operators
\be\label{25}
G_m\ :=\ G_m\!^{\ddt{0}}\quad ,
\ee
the first two of which, $G_0$ and $G_1$, are associated with the
coordinates $X_{0\ad}$ and $X_{1\ad}$. All the operators \gl{25}
share the properties of $G_0$ and $G_1$, e.g. $\{G_m,G_n\}=0$.
Moreover, any two consecutive
operators $G_m$ and $G_{m+1}$ can be taken instead of $G_0$ and $G_1$,
since the choice of four functions from $\{X^{m\ad}\}$
as four target coordinates is arbitrary.
For this reason, the string field $\Upsilon_{+-}$ (depending now
on $X^{m\ad}$) from \gl{19} may be subjected to the conditions
(cf.\cite{IvLe})
\be\label{26}
Q_m\Upsilon_{+-}\ :=\ (G_{m+1}+\zeta G_{m})\Upsilon_{+-}\ =\ 0
\qquad {\rm with} \quad m=0,\ldots,2J{-}1 \quad.
\ee
String fields annihilated by the operators $Q_m$
will be called chiral fields.
As before, $\Upsilon_{+-}$ is a free chiral bosonic string field
with hidden Chan-Paton factors.

\medskip

Substituting the splitting \gl{19} into \gl{26} and using an extension
to Liouville's theorem, we obtain
\be\label{27}
\Psi_+(G_{m+1}+\zeta G_{m})\Psi_+^{-1}\ =\
\Psi_-(G_{m+1}+\zeta G_{m})\Psi_-^{-1}\ =\
A_{m+1}+\zeta\tilde A_m\quad ,
\ee
where $A_{m+1}$ and $\tilde A_m$ are some string fields not depending on
$\zeta$.
Equations \gl{27} can be rewritten as linear equations on $\Psi_\pm$,
\be\label{28}
(G_{m+1}+A_{m+1}+\zeta G_{m} +\zeta \tilde A_{m})\Psi_\pm\ =\ 0\quad .
\ee
The compatibility conditions of Eqs.\gl{28} are the following equations
\bea \nonumber
0 &=& G_{m+1}A_{n+1} + G_{n+1}A_{m+1}+ A_{m+1}A_{n+1} +
A_{n+1}A_{m+1} \quad, \\[8pt]
\label{29}
0 &=& G_m\tilde A_n+G_n\tilde A_m+ \tilde A_m\tilde A_n +
\tilde A_n\tilde A_m  \quad, \\[8pt]
\nonumber
0 &=& G_mA_{n+1}+G_nA_{m+1}+ G_{n+1} \tilde A_m+ G_{m+1} \tilde A_n +
\{\tilde A_m, A_{n+1}\} +\{\tilde A_n, A_{m+1}\} \quad,
\eea
where $m,n=0,\ldots,2J{-}1$.
We name Eqs.\gl{29} the {\it truncated BS hierarchy\/} equations.
For $J\to\infty$ the {\it full BS hierarchy\/} equations emerge.

\medskip

Imposing  on $\Psi_\pm$  the asymptotic conditions \gl{17} and \gl{18},
where $\Phi$ and $\Psi$ depend now on all $X^{m\ad}$, we obtain the gauges
\bea\label{30}
\tilde A_m&=&0 \qquad{\rm and}\qquad
       A_{m+1}\ =\ e^{-\Phi}G_{m+1}e^{\Phi} \quad, \\[6pt]
\label{31}
\tilde A_m&=&0 \qquad{\rm and}\qquad
       A_{m+1}\ =\ G_{m}{\Psi} \quad.
\eea
For the Leznov gauge \gl{31} the BS hierarchy \gl{29} reduces
to the following equations:
\be\label{32}
G_{m+1}G_n\Psi+G_{n+1}G_m\Psi+G_{m}\Psi G_n\Psi+G_{n}\Psi G_m\Psi\ =\ 0\quad,
\ee
containing the Leznov-type Eq.\gl{6} at $m=n=0$.

\medskip

The first novel equations beyond \gl{6} appear for $m=0, n\ge1$.
Acting by $G_0$ on the left yields
\be\label{33}
G_{1}G_0(G_n\Psi )+ G_{0}\Psi G_0(G_n\Psi )
- G_{0}(G_n\Psi )G_0\Psi\ =\ 0\quad .
\ee
Comparing this with \gl{21}, we discover that
\be\label{34}
\de_n\Psi\ =\ G_n\Psi
\ee
are solutions of Eqs.\gl{21}.
Moreover, Eq.\gl{32} provides a recursion relation,
\be\label{35}
\de_{n+1}\Psi\ \equiv\ G_{n+1}\Psi\ =\ G^{-1}_0(G_1\de_{n}\Psi +
G_{0}\Psi \de_{n}\Psi + \de_{n}\Psi G_{0}\Psi )\quad ,
\ee
from which we see that the
symmetries \gl{34} are {\it non-local\/} if $n\ne 0,1$.
It is important to note that $G_0$ has a non-zero kernel,
so that \gl{6} or \gl{35} for $n{=}0$ does not imply
$G_1\Psi$ being $O(\Psi^2)$.
Analogously
one can describe the symmetries of the Yang-type equation \gl{5}.

\medskip

On the {\it linearized level\/} and on mass-shell
the string field $\Psi$ reduces to a first-quantized vertex operator~\cite{BS}.
Up to gauge transformations, the only momentum-dependent $U(1)$-neutral
vertex operator satisfying the linear ``wave'' equation $G_1G_0\Psi=0$
is (see \cite{BV})
\be\label{38}
\Psi\ =\ A\,\exp (ik_{0\dt\a}x^{0\dt\a} + ik_{1\dt\a}x^{1\dt\a})\quad ,
\ee
where the momenta $k_{0\dt\a}$ and $k_{1\dt\a}$ satisfy the on-shell condition
\be\label{36}
k_{0\dt 0}k_{1\dt 1}- k_{0\dt 1}k_{1\dt 0}\ =\ 0 \quad .
\ee
Here, $A$ is a multiplier depending on moduli parameters
$x^{m\dt\a}$ with $m{\ge}2$ but not on space-time coordinates
$x^{0\dt\a}$ or $x^{1\dt\a}$.
Moreover, on the linearized level the action of the
operators $G_0$ and $G_1$ corresponds merely to a multiplication of the vertex
operator~$\Psi$ by $\psi^{\dt\a\ddz}k_{0\dt\a}$ and  $\psi^{\dt\a\ddz}
k_{1\dt\a}$, respectively \cite{BV}.
Thanks to \gl{36}, this implies the identity
\be\label{40}
G_0\Psi\ =\ h(k)\, G_1\Psi
\ee
where \cite{BV}
\be\label{41}
h(k)\ :=\ \frac{k_{0\dt0}}{k_{1\dt0}}\ =\ \frac{k_{0\dt1}}{k_{1\dt1}}\quad.
\ee

\medskip

Being recursively connected with space-time translations,
our symmetries \gl{35} can be compared with the hidden symmetries
of the open $N{=}2$ string constructed
in \cite{IvLe} from momentum operators $P_{0\dt\a}$ plus operators of
picture-raising and spectral flow in the first-quantized BRST approach.
The linearization of the BS hierarchy in Leznov gauge \gl{32} reads
\be\label{37}
G_{m+1}G_n\Psi + G_{n+1}G_m\Psi\ =\ 0\quad .
\ee
Taking $m{=}0$ or linearizing \gl{35} yields
\be\label{39}
G_n\Psi\ =\ G_0^{-1}(G_1 G_{n-1}\Psi)
\ee
which, due to \gl{40}, simplifies to
\be\label{42}
\delta_n^{{\rm lin}}\Psi\ =\ G_n\Psi\ =\ h(k)^{-1}\,G_{n-1}\Psi\
=\ h(k)^{-n}\,G_0\Psi\ =\ h(k)^{-n} k_{0\dt\a}\psi^{\ad\ddz}\Psi \quad .
\ee
It follows that Eqs.\gl{37} for $m,n\ge 1$ are automatically satisfied.
With $\delta_n^{{\rm lin}}=\psi^{\ad\ddz}\delta_{n\ad}^{{\rm lin}}$,
formulae \gl{42} are identical to formulae (5.15) of Ref.\cite{IvLe}.
Thus, we have described an infinite set of non-local symmetries
of the covariant field theory for open $N{=}2$ strings and shown
that the linear part \gl{42} of these symmetries coincides with the
hidden string symmetries \cite{IvLe} found in the BRST approach.
Recall that those symmetries are related to flows on the moduli
space of self-dual gauge fields.

\medskip

It has been demonstrated in the topological approach \cite{BV}
that Eq.\gl{40} implies the (on-shell) vanishing of any $N{=}2$ string
$n$-point amplitude unless
\be
h(k_1)\ =\ h(k_2)\ =\ \ldots\ =\ h(k_n)
\ee for the external momenta $k_1,k_2,\ldots,k_n$,
which is equivalent to all Mandelstam variables being zero.
The same result has been obtained in the BRST approach \cite{JLP,JL2}
by employing the Ward identities associated with the linearized
symmetries~\gl{42}.\footnote{
These facts refer to the closed string but extend to the open case as well
\cite{IvLe}. }
Invoking $S$-matrix analyticity in the Mandelstam variables,
one concludes that only the three-point function escapes the vanishing
theorem.~\footnote{
There is no conflict with the Coleman-Mandula theorem since the unitarity
argument that would kill also the three-point function does not apply
in $2{+}2$ dimensions.}
To sum up, we have established the integrable BS hierarchy describing the
expected higher-spin symmetries as being responsible for the vanishing
results of $N{=}2$ open string scattering \cite{BV}.

\vskip2.cm

\noindent
{\bf\large Acknowledgements}\\[6pt]
This work was partially supported by the German Science Foundation (DFG)
under the grant LE 838/7-1 and the DFG-RFBR grant no. 99-02-04022.  
A.D.P. acknowledges support by RFBR under the grant no. 99-01-01076.

\end{document}